\documentclass[12pt,draftcls,onecolumn]{IEEEtran}
\ifCLASSINFOpdf
  \usepackage[pdftex]{graphicx}
  \graphicspath{{../pdf/}{../jpeg/}}
  \DeclareGraphicsExtensions{.pdf,.jpeg,.png}
\else
  \usepackage[dvips]{graphicx}
  \graphicspath{{../eps/}}
  \DeclareGraphicsExtensions{.pdf}
\fi
\usepackage{epstopdf}
\usepackage{stfloats}
\usepackage[cmex10]{amsmath}
\usepackage{algorithmic}
\usepackage{array}
\usepackage{mdwmath}
\usepackage{mdwtab}
\usepackage{graphicx}
\usepackage{subfigure}
\usepackage{setspace}
\usepackage{amssymb,amsmath,amsthm,amsfonts}
\usepackage{color}
\usepackage{hyperref} 

\begin{document}

\title{Secrecy Outage Analysis of Mixed RF-FSO Downlink SWIPT Systems\thanks{Manuscript received.}}

\author{Hongjiang~Lei,
        Zhijun~Dai,
        Ki-Hong~Park,\\
        Weijia~Lei,
        Gaofeng~Pan,
        and~Mohamed-Slim~Alouini
\thanks{H. Lei, Z. Dai and W. Lei are with Chongqing Key Lab of Mobile Communications Technology, Chongqing University of Posts and Telecommunications, Chongqing 400065, China. H. Lei is also with CEMSE Division, King Abdullah University of Science and Technology (KAUST), Thuwal 23955-6900, Saudi Arabia (e-mail: leihj@cqupt.edu.cn; cquptdzj@gmail.com; leiwj@cqupt.edu.cn).}
\thanks{G. Pan is with Chongqing Key Laboratory of Nonlinear Circuits and Intelligent Information Processing, Southwest University, Chongqing 400715, (e-mail: penngaofeng@qq.com).}
\thanks{K.-H. Park and M.-S. Alouini are with CEMSE Division, King Abdullah University of Science and Technology (KAUST), Thuwal 23955-6900, Saudi Arabia (e-mail: kihong.park@kaust.edu.sa; slim.alouini@kaust.edu.sa).}
}

\maketitle

\begin{abstract}
We analyze a secure dual-hop mixed radio frequency-free space optical (RF-FSO) downlink simultaneous wireless information and power transfer (SWIPT) systems. The FSO link and all RF links experience Gamma-Gamma, independent and identical Nakagami-$m$ fading, respectively. We analyze the effects of atmospheric turbulence, pointing error, detection technology, path loss, and energy harvesting on secrecy performance. Signal-to-noise ratios at both legitimate and illegitimate receivers are not independent since they are both simultaneously influenced by the FSO link. We derive the closed-form expression of the secrecy outage probability (SOP) as well as the asymptotic result for SOP when signal-to-noise ratios at relay and legitimate destinations tend to infinity. \textcolor[rgb]{0.00,0.07,1.00}{Monte-Carlo simulations are performed to verify the accuracy of our analysis. The results show that the secrecy diversity order (SDO) depends on the fading parameter of the relay-destination link and the number of the destination's antennas. Additionally, the SDO also depends on the fading parameters, the pointing error parameter, and the detection type of the FSO link.}
\end{abstract}



\section{Introduction}
\subsection{Background and Related Works}
Dual-hop mixed radio frequency-free space optical (RF-FSO) systems are designed to overcome atmospheric turbulence and other factors limiting the applications of FSO systems. They can also effectively improve communication coverage, save spectrum resources, avoid relocating devices, and are considered as a powerful candidate for next generation of wireless communications \cite{YangL2014JPL, Yang2015JSAC, Yang2017TCOM}. %
In a typical mixed RF-FSO system, users' signals are transmitted to the base station (which serves as a relay node) via the RF link, converted to optical signals, multiplexed, and transmitted to the data center via the FSO link. We called this ``uplink (UL) scenarios''. On the contrary, in ``downlink (DL) scenarios'', messages sent from the data center are delivered to the base station/relay through the FSO link, converted to a wireless signal and then sent to the user.
Assuming a message sent by the data center is only for specific users, the remaining users within the coverage of the relay node are potential eavesdroppers.

It has been verified that physical layer security (PLS) technology can prevent illegitimate receivers from eavesdropping due to the time-varying nature of the wireless medium \cite{Zou2016Proc}-\cite{Lei2016SPL}.
Numerical studies of PLS over FSO satellite ground systems were performed by Endo \emph{et al.} \cite{Endo2015PJ}. They showed that secrecy communications were possible and that there can be a complementary technologies to balance security and usability issues. But in their study, Endo \emph{et al.} only considered some idealistic conditions and assumed that the channels were fading-free. Lopez-Martinez \emph{et al.} \cite{Lopez2015PJ} studied PLS based on Wyner's FSO model and used the probability of strict secrecy capacity to evaluate the secrecy performance. But they considered only two special cases: when the eavesdropper is either near the source or the destination. Sun and Djordievic \cite{Sun2016PJ} studied a secure orbital angular momentum multiplexing FSO system and numerically simulated its secrecy capacity. Their results showed that secrecy performance depends on the location of eavesdroppers and that orbital angular momentum multiplexing technology could improve the secrecy in weak and medium turbulence regimes.

The FSO link is viewed to be highly secure since the laser beam has high directionality \cite{Malek2016TWC}-\cite{Leihj2018PJ}. However, the broadcast nature of the RF link makes the mixed RF-FSO systems vulnerable to wiretap. Recently, the PLS of mixed RF-FSO systems stimulated researchers' interest and quite a few studies on this topic were reported in the literature, including \cite{Malek2016TWC}-\cite{Leihj2018PJ}.
El-Malek \emph{et al.} \cite{Malek2016TWC} studied the security reliability trade-off of a single-input multiple-output mixed RF-FSO system and derived the closed-form expressions for some generalized performance metrics, such as outage probability (OP), intercept probability (IP), \emph{etc}. The same authors also analyzed the effect of RF co-channel interference on the secrecy performance of mixed RF-FSO systems and proposed a new power allocation scheme to enhance the secrecy performance \cite{Malek2017JLT}.
Notice that the all the performance metrics studied in \cite{Malek2016TWC} and \cite{Malek2017JLT} were considered the main channel or wiretap channel separately. By contrast, the secrecy outage probability (SOP) investigated in our work is considered the main channel and wiretap channel simultaneously.
We studied the secrecy performance of a UL mixed RF-FSO system with perfect and imperfect channel state information in \cite{Leihj2017PJ} and \cite{Leihj2018PJ}, respectively. The closed-formed expressions for the exact and asymptotic SOP were derived. Our results demonstrated that the turbulence degrades the secrecy performance and that it is difficult to wiretap when the intensity modulation with direct detection (IM/DD) technology is replaced by the heterodyne detection (HD) technology.
Furthermore, the secrecy outage performance of a UL mixed RF-FSO system with  was investigated in

It is noteworthy that \cite{Malek2016TWC}-\cite{Leihj2018PJ} considered the UL mixed RF-FSO transmission systems, in which the FSO link only influences the signal-to-noise ratio (SNR) at legitimate receivers.
Technically speaking, it is much more challenging to analyze the secrecy performance for DL mixed RF-FSO transmission systems compared to analyze the OP/IP/SOP for UL mixed RF-FSO transmission systems. This is because the problem in DL mixed RF-FSO systems becomes complex where all RF destinations are affected by the FSO link.

It is expected that the next generation of wireless communications will comprise a lot of simpler and cheaper wireless nodes which are powered by batteries. For these nodes, it is quite difficult or even impossible to replace the batteries. Then the simultaneous wireless information and power transfer (SWIPT) technology was proposed to solve this problem \cite{ZhangR2013TWC}-\cite{Wu2015Mag}.
Since part of energy is used to charge the  battery at receivers, the power for information delivery will decrease, which will lead to the  degraded secrecy capacity \cite{lei2017CL}. Thus many literature recently focused on  the security of SWIPT systems
\cite{DWKNg2014TWC}-\cite{Pan2017TCOMVLC}.
\textcolor[rgb]{0.00,0.07,1.00}{The security for SWIPT systems was first considered in \cite{DWKNg2014TWC} and the resource allocation design for secure MISO SWIPT systems was formulated as a non-convex optimization problem and an efficient resource allocation algorithm was proposed to obtain the global optimal solution.}
The secrecy performance of SIMO and MISO SWIPT systems were investigated and the closed-form expressions for SOP were derived in \cite{Pan2015TCOM} and \cite{Pan2016TCOM}, respectively.
The secrecy outage performance of an underlay multiple-input-multiple-output cognitive radio networks with energy harvesting and transmit antenna selection was studied in \cite{Lei2017TGCN}.
And all these works just considered the RF systems.
Pan \emph{et al.} investigated the secrecy performance of a hybrid visible light communication (VLC)-RF system with light energy harvesting and derived analytical expressions for exact and asymptotic SOP in \cite{Pan2017TCOMVLC}.
Makki \emph{et al.} analyzed the throughput and OP for the hybrid RF-FSO SWIPT systems and a power allocation scheme was proposed in \cite{Makki2017WCL}.
But all these works considered the hybrid systems, in which the FSO/VLC and RF links were parallel and backup/backhaul. In our work, a dual-hop mixed RF-FSO system is considered.
It is assumed that SWIPT is used to collect energy for all RF receivers from the wireless signals sent by the relay node.

\subsection{Motivation and Contributions}
To our best knowledge, there is no literature studying the physical layer security of DL mixed RF-FSO SWIPT systems. In this work, we study a secure DL mixed RF-FSO system and analyze the effects of misalignment, different detection schemes, SWIPT, and multiple antenna techniques on secrecy performance of mixed systems. In summary:

\begin{itemize}
\item We study the secrecy outage performance of the DL mixed RF-FSO SWIPT systems over Gamma-Gamma - Nakagami-$m$ fading channels with DF relaying schemes. We investigate the effects of misalignment, different detection schemes, SWIPT, and multiple antenna techniques and deduct the closed-form expressions for the exact and asymptotic SOPs.
\item We present a selected figures illustrating Monte-Carlo simulations and analytic results in order to validate our analysis. Results show that the HD detection method can lead to lower secrecy outage compared to IM/DD, and that the SOP can also be improved with less pointing error or/and weak turbulence. The path-loss degrades the security of the DL mixed RF-FSO systems when the FSO link is the bottleneck of the transmission, and vice versa. Moreover, our results show that the secrecy diversity order (SDO) is determined by the fading parameter of the relay-destination link and the number of  the destination's antennas. Additionally, the SDO also depends on the fading parameters, the pointing error parameter, and the detection type of the FSO link.
\item \textcolor[rgb]{0.00,0.07,1.00}{The correlation of the SNR at legitimate and illegitimate receivers is considered and is eliminated by using the law of total probability. The results in our work does not only apply to the mixed RF-FSO systems but also can be utilized to investigate the secrecy performance of all the dual-hop cooperative systems with DF scheme when there is not direct link between the source and the receiver.}
\item Differing from \cite{Malek2016TWC}-\cite{Leihj2018PJ}, the DL mixed RF-FSO systems were studied in this work, in which both SNRs at legitimate and illegitimate destinations are influenced by the FSO link.
\item Although the secrecy performance of DL mixed RF-FSO systems was investigated in \cite{Kumar2018IJCS}, the correlation of the SNR at both legitimate and illegitimate receivers was not considered. Furthermore, SWIPT, path-loss fading, and multiple antennas are considered in this work.

\end{itemize}

The rest of the paper is organized as follows. Section II describes the system model. The exact SOP analysis is presented in Section III, while Section IV analyzes the asymptotic SOP. Simulation results are given in Section V, while Section VI concludes this paper.

\section{System Models}
We consider a DL mixed RF-FSO SWIPT system (shown in Fig. 1), with confidential signals transmitted from the data center $\left( S \right)$ to the legitimate destination node $\left( D \right)$ through the relay $\left( R \right)$. There is an eavesdropper $E$ who is attempting to wiretap the information, and $D$ and $E$ are equipped with ${N_D}\left( {{N_D} \ge 1} \right)$ and ${N_E}\left( {{N_E} \ge 1} \right)$ antennas, respectively. We assume that the FSO link follows a unified Gamma-Gamma fading and that all the RF links experience independent and identical Nakagami-$m$ fading. The maximum ratio combining (MRC) scheme is utilized at both $D$ and $E$ to improve the received SNR.

All the receivers (both $D$ and $E$) are equipped with a rechargeable battery harvesting the RF energy broadcasted from $R$, and power splitting (PS) method is used to coordinate the processes of information decoding and energy harvesting from the received signal \cite{ZhangR2013TWC, Pan2015TCOM, Pan2016TCOM}. This means that the received signal is divided into  information decoding (ID) part and harvesting energy (EH) part. In other words, the ${\alpha _j}\left( {0 \le {\alpha _j} \le 1}, j \in \left\{ {D,E} \right\} \right)$ portion of the signal power is used to decode information, and the remaining  portion of power is used for harvesting the energy.
\textcolor[rgb]{0.00,0.07,1.00}{
The linear EH model is not practical since an EH circuit usually comprises diodes, inductors and capacitors. The new non-linear EH model was proposed in \cite{Boshkovska2015CL} and \cite{Dong2016CL}, respectively.
Actually, the EH model does not influence the secrecy performance of the mixed RF-FSO system because the different
EH model just influence the energy harvested at the receivers ($D$ and $E$) and does not influence the SNR at the receivers. Only the portion of signal power used to ID influences the SNR at receivers. The situation is similar to the case with the infinity capacity EH buffer and finite capacity EH buffer scenarios. Our results are also fit to the case with time splitting method. It should be noted that the splitting factors in our work are assumed to be fixed. The secrecy performance might be affected with the splitting factors which are dynamically varying depending on the non-linear EH model, which will be addressed in our future work.
}
\begin{figure}[h]
\centering{\includegraphics[width = 4 in]{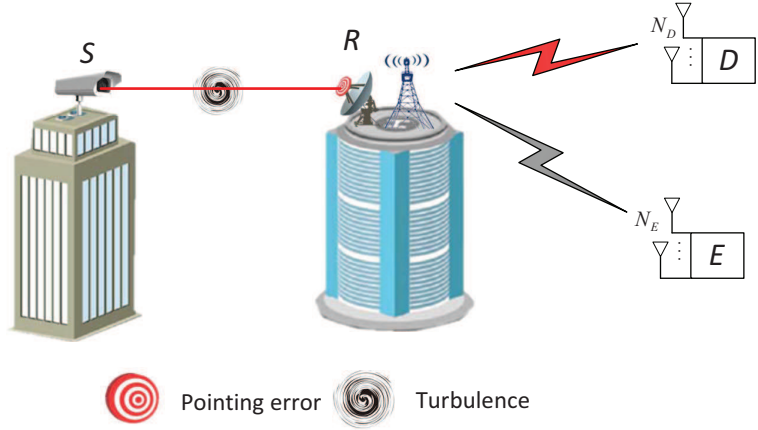}}
\caption{System model of a DL mixed RF-FSO SWIPT system which consists of the data center ($S$), the relay ($R$), the legitimate destination ($D$), and the eavesdropper ($E$). Both $D$ and $E$ are equipped with multiple antennas.}
\end{figure}

The probability density function (PDF) and cumulative distribution function (CDF) of ${\gamma _{SR}}$ can be expressed as \cite{Ansari2015VTC}
\begin{equation}
{f_{{\gamma _{SR}}}}\left( \gamma  \right) = A{\gamma ^{ - 1}}G_{1,3}^{3,0}\left[ {B{\gamma ^{\frac{1}{r}}}\left| {_{{\xi ^2},a ,b }^{{\xi ^2} + 1}} \right.} \right],
\label{pdfsr}
\end{equation}
\begin{equation}
{F_{{\gamma _{SR}}}}\left( \gamma  \right) = I G_{r + 1,3r + 1}^{3r,1}\left[ {\rho \gamma \left| {_{{K_2},0}^{1,{K_1}}} \right.} \right],
\label{cdffso}
\end{equation}
respectively, where $A = \frac{{{\xi ^2}}}{{r\Gamma \left( a  \right)\Gamma \left( b  \right)}}$, $B = \frac{{hab}}{{\sqrt[r]{{{\Omega _{SR}}}}}}$, $I  = \frac{{{\xi ^2}{r^{a + b - 2}}}}{{{{\left( {2\pi } \right)}^{r - 1}}\Gamma \left( a \right)\Gamma \left( b \right)}}$. $a$ and $b$ are the fading parameters, $r$ represents the detection scheme used at $R$, i.e. $r = 1$ for HD and $r = 2$ for IM/DD, $\xi$ is the pointing error at the destination \cite{Zedini2016TWC}.
$\rho  = \frac{{{{\left( {hab} \right)}^r}}}{{{\Omega _{SR}}{r^{2r}}}}$,
${{\Omega _{SR}}}$ represents the electrical SNR of the FSO link,
${K_1} = \Delta \left( {r,{\xi ^2} + 1} \right)$, ${K_2} = \left[ {\Delta \left( {r,{\xi ^2}} \right),\Delta \left( {r,a} \right),\Delta \left( {r,b} \right)} \right]$, $\Delta \left( {k,a} \right) = \frac{a}{k},\frac{{a + 1}}{k}, \cdots ,\frac{{a + k - 1}}{k}$, $h = \frac{{{\xi ^2}}}{{{\xi ^2} + 1}}$, and $G_{p,q}^{m,n}\left[  \cdot  \right]$ is Meijer's $G$-function, as defined by  (9.301) of \cite{Gradshteyn2007}.

The effects of path-loss and small-scale fading of the RF link are considered and the received signals at the $m$th  $\left( {1 \le m \le {N_D}} \right)$ antennas of $D$ are expressed as
\begin{equation}
{y_{D,m}} = \sqrt {{\alpha _D}} \left( {\sqrt {{P_t}{L_c}d_{D}^{ - \eta }} {h_{R{D_m}}}x + {n_{D,m}}} \right) + {z_{D,m}},
\end{equation}
where ${P_t}$ is the transmit power at $R$, ${L_c}$ is the propagation loss constant, ${d_D}$ is the distance between $R$ and $D$, $\eta$ is the path-loss exponent ($\eta = 0$ means that the effect of path loss is ignored), $x$ denotes the transmitted symbol from $R$, ${h_{R{D_m}}}$ is the channel coefficient, ${n_{D,m}}$ and ${z_{D,m}}$ represent the additive white Gaussian noise and the signal processing noise at the $m$th antenna of $D$, which are additive white Gaussian noise with zero means and variances ${N_0}$ and $\sigma _D^2$, respectively.

The SNR of the signal at $D$ is then written as
\begin{equation}
{\gamma _{RD}} = \frac{{{\alpha _D}{P_t}{L_c}\sum\limits_{m = 1}^{{N_D}} {{H_{RD_m}}} }}{{d_D^\eta \left( {{\alpha _D}{N_0} + \sigma _D^2} \right)}},
\end{equation}
where ${H_{RD_m}} = {\left| {{h_{RD_m}}} \right|^2}$.

Using Lemma 1 of \cite{Lei2017TVTTAS}, we obtain the PDF and CDF for the received SNR on $D$ as
\begin{equation}
{f_{{\gamma _{RD}}}}\left( \gamma  \right) = \frac{{\lambda _D^{{\tau _D}}}}{{\Gamma \left( {{\tau _D}} \right)}}{\gamma ^{{\tau _D} - 1}}{e^{ - {\lambda _D}\gamma }},
\label{pdfrd}
\end{equation}
\begin{equation}
{F_{{\gamma _{RD}}}}\left( \gamma  \right) = 1 - {e^{ - {\lambda _D}\gamma }}\sum\limits_{p = 0}^{{\tau _D} - 1} {\frac{{\lambda _D^p}}{{p!}}{\gamma ^p}},
\label{cdfrd}
\end{equation}
respectively, where ${\lambda _D} = \frac{{{m_D}d_D^\eta \left( {{\alpha _D}{N_0} + \sigma _D^2} \right)}}{{{\Omega _{RD}}{\alpha _D}{P_t}{L_c}}}$, ${\Omega _{RD}}$ is the expectation of ${H_{RD_m}}$, ${\tau _D} = {m_D}{N_D}$, and $\Gamma \left(  \cdot  \right)$ is the Gamma function, as defined by eq. (8.310) of \cite{Gradshteyn2007}.

Similarly, we obtain the PDF and CDF of ${\gamma _{RE}}$ as
\begin{equation}
{f_{{\gamma _{RE}}}}\left( \gamma  \right) = \frac{{\lambda _E^{{\tau _E}}}}{{\Gamma \left( {{\tau _E}} \right)}}{\gamma ^{{\tau _E} - 1}}{e^{ - {\lambda _E}\gamma }},
\label{pdfre}
\end{equation}
\begin{equation}
{F_{{\gamma _{RE}}}}\left( \gamma  \right) = 1 - {e^{ - {\lambda _E}\gamma }}\sum\limits_{n = 0}^{{\tau _E} - 1} {\frac{{\lambda _E^n}}{{n!}}{\gamma ^n}},
\label{cdfre}
\end{equation}
respectively, where ${\lambda _E} = \frac{{{m_E}d_E^\eta \left( {{\alpha _E}{N_0} + \sigma _E^2} \right)}}{{{\Omega _{RE}}{\alpha _E}{P_t}{L_c}}}$, ${\Omega _{RE}}$ is the average power channel gains between $R$ and $E$, and ${\tau _E} = {m_E}{N_E}$.

We assume that \textcolor[rgb]{0.00,0.07,1.00}{decode-and-forward (DF) relaying scheme }is used at $R$. The equivalent SNRs at $D$ and $E$ are then expressed as \footnote{\textcolor[rgb]{0.00,0.07,1.00}{The results are also fit to the bound of variable gain amplify-and-forward relaying scheme, as testified in many literature, such as \cite{Leihj2017PJ, Zedini2016TWC, Zedini2015PJ}.} }
\begin{equation}
{\gamma _{eq,D}} = \min \left( {{\gamma _{SR}},{\gamma _{RD}}} \right)
\label{snrsrdeq}
\end{equation}
\begin{equation}
{\gamma _{eq,E}}  = \min \left( {{\gamma _{SR}},{\gamma _{RE}}} \right),
\label{snrsreeq}
\end{equation}
respectively.

\section{Secrecy Outage Probability Analysis}
As defined in \cite{Bloch2008}, we obtain the secrecy capacity of DF relaying scheme as
\begin{equation}
\begin{aligned}
{C_s} &= \max \left\{ {{C_{eq,D}} - {C_{eq,E}},0} \right\}\\
&= \max \left\{ {\ln \left( {1 + \min \left( {{\gamma _{SR}},{\gamma _{RD}}} \right)} \right) - \ln \left( {1 + \min \left( {{\gamma _{SR}},{\gamma _{RE}}} \right)} \right),0} \right\}\\
&= \left\{ {\begin{array}{*{20}{c}}
{\max \left\{ {\ln \left( {1 + {\gamma _{SR}}} \right) - \ln \left( {1 + {\gamma _{SR}}} \right),0} \right\},\,\,\,{\gamma _{SR}} \le \min \left( {{\gamma _{RD}},{\gamma _{RE}}} \right)}\\
{\max \left\{ {\ln \left( {1 + {\gamma _{RD}}} \right) - \ln \left( {1 + {\gamma _{RE}}} \right),0} \right\},{\gamma _{SR}} \ge \max \left( {{\gamma _{RD}},{\gamma _{RE}}} \right)}\\
{\max \left\{ {\ln \left( {1 + {\gamma _{SR}}} \right) - \ln \left( {1 + {\gamma _{RE}}} \right),0} \right\},\,\,\,\,\,\,\,\,\,\,\,\,{\gamma _{RE}} \le {\gamma _{SR}} \le {\gamma _{RD}}}\\
{\max \left\{ {\ln \left( {1 + {\gamma _{RD}}} \right) - \ln \left( {1 + {\gamma _{SR}}} \right),0} \right\},\,\,\,\,\,\,\,\,\,\,\,\,{\gamma _{RD}} \le {\gamma _{SR}} \le {\gamma _{RE}}}
\end{array}} \right.\\
& = \left\{ {\begin{array}{*{20}{c}}
{\ln \left( {1 + {\gamma _{RD}}} \right) - \ln \left( {1 + {\gamma _{RE}}} \right),{\gamma _{RE}} \le {\gamma _{RD}} \le {\gamma _{SR}}}\\
{\ln \left( {1 + {\gamma _{SR}}} \right) - \ln \left( {1 + {\gamma _{RE}}} \right),{\gamma _{RE}} \le {\gamma _{SR}} \le {\gamma _{RD}}}\\
{0,\quad \quad \quad \quad \quad \quad \quad \quad \quad {\rm{otherwise}}}
\end{array}} \right.
\end{aligned}
\label{secrecycapacity}
\end{equation}

\textcolor[rgb]{0.00,0.07,1.00}{\emph{\textbf{Remark 1:}} From (\ref{secrecycapacity}), one can easily find that when ${\gamma _{RE}} < {\gamma _{SR}}$, the  secrecy capacity of mixed RF-FSO systems with DF scheme can be rewritten as
\begin{equation}
{C_s} = \max \left\{ {\ln \left( {1 + \min \left( {{\gamma _{SR}},{\gamma _{RD}}} \right)} \right) - \ln \left( {1 + {\gamma _{RE}}} \right),0} \right\}.
\label{Cs2}
\end{equation}
Eq. (\ref{Cs2}) means that the $R-E$ link is the bottleneck for the equivalent SNR at $E$. This equation is easy to understand but very useful since the two parts in (\ref{Cs2}) are independent. It should be noted that the secrecy capacity in this case (when ${\gamma _{RE}} < {\gamma _{SR}}$) may be equal to zero.}

\textcolor[rgb]{0.00,0.07,1.00}{{\emph{\textbf{Remark 2:}}} On the other side, when ${\gamma _{RE}} > {\gamma _{SR}}$, which means $S$-$R$ link is the bottleneck for the equivalent SNR at $E$, the secrecy capacity in this case must be zero because the equivalent SNR at $D$ cannot be greater than the one at $R$, which can be expressed as
\begin{equation}
{\min \left( {{\gamma _{SR}},{\gamma _{RD}}} \right) \le \min \left( {{\gamma _{SR}},{\gamma _{RE}}} \right) = {\gamma _{SR}}} .
\end{equation}
}

Thus SOP can be expressed as
\begin{equation}
\begin{aligned}
{P_{out}} &= \Pr \left\{ {{C_s} \le {R_s}} \right\}\\
& = {H_1} + {H_2} + 1 - \varrho,
\end{aligned}
\label{pout}
\end{equation}
where $R_s$ represents the target secrecy rate, $H_1$, $H_2$, and $\varrho$ are expressed as
\begin{equation}
{H_1} = \Pr \left\{ {\ln \left( {1 + {\gamma _{RD}}} \right) - \ln \left( {1 + {\gamma _{RE}}} \right) < {R_s},{\gamma _{RE}} \le {\gamma _{RD}} \le {\gamma _{SR}}} \right\}
\label{H10}
\end{equation}
\begin{equation}
{H_2} = \Pr \left\{ {\ln \left( {1 + {\gamma _{SR}}} \right) - \\ \ln \left( {1 + {\gamma _{RE}}} \right) \le {R_s},{\gamma _{RE}} \le {\gamma _{SR}} \le {\gamma _{RD}}} \right\}
\label{H20}
\end{equation}
\begin{equation}
\begin{aligned}
\varrho &= \Pr \left\{ {{\gamma _{RE}} \le {\gamma _{RD}} \le {\gamma _{SR}}} \right\} + \Pr \left\{ {{\gamma _{RE}} \le {\gamma _{SR}} \le {\gamma _{RD}}} \right\}\\
& = \Pr \left\{ {{\gamma _{RE}} \le {\gamma _{eq,D}} = \min \left( {{\gamma _{RD}},{\gamma _{SR}}} \right)} \right\}.
\label{P10}
\end{aligned}
\end{equation}

\textcolor[rgb]{0.00,0.07,1.00}{\emph{\textbf{Remark 3:}} Based on (\ref{H10}) and (\ref{H20}),  one can easily find that $H_1$ and $H_2$ means that the bottleneck of equivalent SNR at $D$ is $R-D$ and $S-R$ link, respectively. The corresponding secrecy capacity in these two cases is positive but less than $R_s$, which cause the secrecy outage.}

\textcolor[rgb]{0.00,0.07,1.00}{\emph{\textbf{Remark 4:}} Moreover, one can find that  (\ref{P10}) has no relationship with $R_s$. Because of ${\gamma _{eq,E}} = \min \left( {{\gamma _{SR}},{\gamma _{RE}}} \right) \le {\gamma _{RE}}$, based on (\ref{P10}), we can observe that $\varrho $ denotes the probability of  ${\gamma _{eq,E}} \le {\gamma _{eq,D}}$. Then $1 - \varrho$ signifies the probability for ${C_s} = 0$ since the secrecy capacity of mixed RF-FSO systems equals zero  when  ${\gamma _{eq,E}} > {\gamma _{eq,D}}$.}

\textcolor[rgb]{0.00,0.07,1.00}{We can easily obtain another useful secrecy metric, probability of strictly positive secrecy capacity \cite{Pan2016TVT, Lei2016CL}, as
\begin{equation}
{P_0} = \varrho - \left( {{H_1} + {H_2}} \right)\left| {_{{R_s} = 0}} \right..
\end{equation}
}

\textcolor[rgb]{0.00,0.07,1.00}{It should also note that the previous results (Eqs. (\ref{secrecycapacity})- (\ref{P10}), \emph{\textbf{Remark}} 1 - 4) does not only apply to the mixed RF-FSO systems but also can be utilized to investigate the secrecy performance of all the dual-hop cooperative systems with DF scheme when there is not direct link between the source and the receiver. }

In the following, we derive the closed-form expressions of (\ref{H10})-(\ref{P10}).

\subsection{Derivations of $H_1$ }
We can rewrite $H_1$ as
\begin{equation}
{H_1} = \Pr \left\{ {{\gamma _{RE}} > \frac{{{\gamma _{RD}} + 1 - \Theta }}{\Theta },{\gamma _{RE}} \le {\gamma _{RD}} \le {\gamma _{SR}}} \right\},
\end{equation}
where $\Theta  = {e^{{R_s}}} \ge 1$.

Since $\frac{{{\gamma _{RD}} + 1 - \Theta }}{\Theta } \le {\gamma _{RD}}$, $H_1$ can be rewritten as
\begin{equation}
\begin{aligned}
{H_1} &= \int_0^\infty  {\int_0^x {{\phi _1}\left( y \right){f_{{\gamma _{RD}}}}\left( y \right)dy} {f_{{\gamma _{SR}}}}\left( x \right)dx} \\
& = \int_0^{\Theta  - 1} {\int_0^x {{F_{{\gamma _{RE}}}}\left( y \right){f_{{\gamma _{RD}}}}\left( y \right)dy} {f_{{\gamma _{SR}}}}\left( x \right)dx} \\
& + \int_{\Theta  - 1}^\infty  {\int_0^x {{\phi _2}\left( y \right){f_{{\gamma _{RD}}}}\left( y \right)dy} {f_{{\gamma _{SR}}}}\left( x \right)dx} \\
& = {H_{11}} + {H_{12}} + {H_{13}},
\end{aligned}
\label{H1}
\end{equation}
where
\begin{align*}
&{H_{11}} = \int_0^{\Theta  - 1} {\int_0^x {{F_{{\gamma _{RE}}}}\left( y \right){f_{{\gamma _{RD}}}}\left( y \right)dy} {f_{{\gamma _{SR}}}}\left( x \right)dx},\\
&{H_{12}} = \int_{\Theta  - 1}^\infty  {\int_0^{\Theta  - 1} {{F_{{\gamma _{RE}}}}\left( y \right){f_{{\gamma _{RD}}}}\left( y \right)dy} {f_{{\gamma _{SR}}}}\left( x \right)dx},\\
&{H_{13}} = \int_{\Theta  - 1}^\infty  {\int_{\Theta  - 1}^x {{\phi _3}\left( y \right){f_{{\gamma _{RD}}}}\left( y \right)dy} {f_{{\gamma _{SR}}}}\left( x \right)dx},\\
&{\phi _1}\left( y \right) = \Pr \left\{ {\frac{{y + 1 - \Theta }}{\Theta } < {\gamma _{RE}} < y} \right\},\\
&{\phi _2}\left( y \right) = {F_{{\gamma _{RE}}}}\left( y \right) - {F_{{\gamma _{RE}}}}\left( {\max \left\{ {0,\frac{{y + 1 - \Theta }}{\Theta }} \right\}} \right),\\
&{\phi _3}\left( y \right) = {F_{{\gamma _{RE}}}}\left( y \right) - {F_{{\gamma _{RE}}}}\left( {\frac{{y + 1 - \Theta }}{\Theta }} \right).
\end{align*}

By substituting (\ref{pdfsr}), (\ref{pdfrd}) and (\ref{cdfre}) into $H_{11}$, we obtain
\begin{equation}
\begin{aligned}
{H_{11}} &= \frac{{{G_0}\left( {{\tau _D},{\lambda _D}} \right)}}{{\Gamma \left( {{\tau _D}} \right)}} - \frac{{\lambda _D^{{\tau _D}}}}{{\Gamma \left( {{\tau _D}} \right)}}\sum\limits_{n = 0}^{{\tau _E} - 1} {\frac{{\lambda _E^n{G_0}\left( {{\tau _D} + n,{\lambda _D} + {\lambda _E}} \right)}}{{n!{{\left( {{\lambda _D} + {\lambda _E}} \right)}^{{\tau _D} + n}}}}},
\end{aligned}
\end{equation}
where $G_0 \left( {\alpha ,\beta } \right)  = \int_0^{\Theta  - 1} {\Upsilon \left( {\alpha ,\beta x} \right){f_{{\gamma _{SR}}}}\left( x \right)dx}$
and
$\Upsilon \left( { \cdot ,  \cdot } \right)$ is the lower incomplete Gamma function, defined by (8.350.1) of \cite{Gradshteyn2007}. The closed-form expression of $G_0 \left( {\alpha ,\beta } \right)$ is given in Appendix A.

By substituting (\ref{pdfsr}), (\ref{pdfrd}) and (\ref{cdfre}) into $H_{12}$ and using (3.351.1) of \cite{Gradshteyn2007}, we obtain
\begin{equation}
\begin{aligned}
{H_{12}} &= \int_{\Theta  - 1}^\infty  {\int_0^{\Theta  - 1} {{F_{{\gamma _{RE}}}}\left( y \right){f_{{\gamma _{RD}}}}\left( y \right)dy} {f_{{\gamma _{SR}}}}\left( x \right)dx} \\
& = \int_{\Theta  - 1}^\infty  {{f_{{\gamma _{SR}}}}\left( x \right)dx} \int_0^{\Theta  - 1} {{F_{{\gamma _{RE}}}}\left( y \right){f_{{\gamma _{RD}}}}\left( y \right)dy} \\
& = \left( {1 - {F_{{\gamma _{SR}}}}\left( {\Theta  - 1} \right)} \right)\left( {\int_0^{\Theta  - 1} {{f_{{\gamma _{RD}}}}\left( y \right)dy}  - \sum\limits_{n = 0}^{{\tau _E} - 1} {\frac{{{\lambda _E}^n}}{{n!}}} \int_0^{\Theta  - 1} {{y^n}{e^{ - {\lambda _E}y}}{f_{{\gamma _{RD}}}}\left( y \right)dy} } \right)\\
& = \left( {1 - {F_{{\gamma _{SR}}}}\left( {\Theta  - 1} \right)} \right)\left( {{F_{{\gamma _{RD}}}}\left( {\Theta  - 1} \right)} - {\frac{{\lambda _D^{{\tau _D}}}}{{\Gamma \left( {{\tau _D}} \right)}}\sum\limits_{n = 0}^{{\tau _E} - 1} {\frac{{{\lambda _E}^n\Upsilon \left( {{\tau _D} + n,\left( {{\lambda _D} + {\lambda _E}} \right)\left( {\Theta  - 1} \right)} \right)}}{{n!{{\left( {{\lambda _D} + {\lambda _E}} \right)}^{{\tau _D} + n}}}}} } \right).
\end{aligned}
\end{equation}

Using of (1.111) and (3.351.1) of \cite{Gradshteyn2007}, we have $H_{13}$ as
\begin{equation}
\begin{aligned}
{H_{13}} & = \int_{\Theta  - 1}^\infty  {\int_{\Theta  - 1}^x {{\phi _3}\left( y \right){f_{{\gamma _{RD}}}}\left( y \right)dy} {f_{{\gamma _{SR}}}}\left( x \right)dx}  \\
& =  \frac{{\lambda _D^{{\tau _D}}{e^{ - \frac{{{\lambda _E}\left( {1 - \Theta } \right)}}{\Theta }}}}}{{\Gamma \left( {{\tau _D}} \right)}} \sum\limits_{n = 0}^{{\tau _E} - 1} {\sum\limits_{t = 0}^n {\frac{{\lambda _E^n{{\left( {1 - \Theta } \right)}^{n - t}}{H_{131}}}}{{t!\left( {n - t} \right)!{\Theta ^n}}}{{\left( {{\lambda _D} + \frac{{{\lambda _E}}}{\Theta }} \right)}^{ - {\tau _D} - t}}} }\\
& - \frac{{\lambda _D^{{\tau _D}}}}{{\Gamma \left( {{\tau _D}} \right)}}\sum\limits_{n = 0}^{{\tau _E} - 1} {\frac{{\lambda _E^n{H_{132}}}}{{n!{{\left( {{\lambda _D} + {\lambda _E}} \right)}^{{\tau _D} + n}}}}} ,
\end{aligned}
\end{equation}
where
${H_{131}} = G_2 \left( {{\tau _D} + t,\left( {{\lambda _D} + \frac{{{\lambda _E}}}{\Theta }} \right)} \right) -
 \Upsilon \left( {{\tau _D} + t,\left( {{\lambda _D} + \frac{{{\lambda _E}}}{\Theta }} \right)\left( {\Theta  - 1} \right)} \right)\left( {1 - {F_{{\gamma _{SR}}}}\left( {\Theta  - 1} \right)} \right)$,
${H_{132}} = G_2 \left( {{\tau _D} + n,\left( {{\lambda _D} + {\lambda _E}} \right)} \right)
- \Upsilon \left( {{\tau _D} + n,\left( {{\lambda _D} + {\lambda _E}} \right)\left( {\Theta  - 1} \right)} \right)\left( {1 - {F_{{\gamma _{SR}}}}\left( {\Theta  - 1} \right)} \right)$, and
$G_2 \left( {\alpha ,\beta } \right) = A \int_{\Theta  - 1}^\infty  {{x^{ - 1}}\Upsilon \left( {\alpha ,\beta x} \right)G_{1,3}^{3,0}\left[ {B{x^{\frac{1}{r}}}\left| {_{{\xi ^2},a,b}^{{\xi ^2} + 1}} \right.} \right]dx}$.
The closed-form expression of $G_2 \left( {\alpha ,\beta } \right)$ is given in Appendix B.

\subsection{Derivations of $H_2$}
Similar to the derivation of $H_1$, we have
\begin{equation}
{H_2} = {H_{21}} + {H_{22}} + {H_{23}},
\label{H2}
\end{equation}
where ${H_{21}} = \int_0^{\Theta  - 1} {\int_0^x {{F_{{\gamma _{RE}}}}\left( y \right){f_{{\gamma _{SR}}}}\left( y \right)dy} {f_{{\gamma _{RD}}}}\left( x \right)dx} $,
${H_{22}} = \int_{\Theta  - 1}^\infty  {\int_0^{\Theta  - 1} {{F_{{\gamma _{RE}}}}\left( y \right){f_{{\gamma _{SR}}}}\left( y \right)dy} {f_{{\gamma _{RD}}}}\left( x \right)dx} $, and
${H_{23}} = \int_{\Theta  - 1}^\infty  {\int_{\Theta  - 1}^x {{\phi _3}\left( y \right){f_{{\gamma _{SR}}}}\left( y \right)dy} {f_{{\gamma _{RD}}}}\left( x \right)dx}$.

Substituting (\ref{pdfsr}), (\ref{pdfrd}) and (\ref{cdfre}) into $H_{21}$ and using (07.34.21.0084.01) of \cite{Wolfram2001}, we can achieve
\begin{equation}
\begin{aligned}
{H_{21}} &= \int_0^{\Theta  - 1} {{\int_0^x {{F_{{\gamma _{RE}}}}\left( y \right){f_{{\gamma _{SR}}}}\left( y \right)dy} } {f_{{\gamma _{RD}}}}\left( x \right)dx} \\
& = \frac{{I\lambda _D^{{\tau _D}}}}{{\Gamma \left( {{\tau _D}} \right)}}\int_0^{\Theta  - 1} {{x^{\alpha  - 1}}{e^{ - \beta x}}G_{r + 1,3r + 1}^{3r,1}\left[ {\rho x\left| {_{{K_2},0}^{1,{K_1}}} \right.} \right]dx} \\
&  - \frac{{A\lambda _D^{{\tau _D}}}}{{\Gamma \left( {{\tau _D}} \right)}}\sum\limits_{n = 0}^{{\tau _E} - 1} {\frac{{\lambda _E^n}}{{n!}}\int_0^{\Theta  - 1} {\int_0^x {{y^{n - 1}}} } e^{ - {\lambda _E}y}}G_{1,3}^{3,0}\left[ {B{y^{\frac{1}{r}}}\left| {_{{\xi ^2},a,b}^{{\xi ^2} + 1}} \right.} \right]{x^{{\tau _D} - 1}}{e^{ - {\lambda _D}x}}dydx \\
& = \frac{I}{{\Gamma \left( {{\tau _D}} \right)}}\sum\limits_{s = 0}^\infty  {\frac{{{{\left( { - 1} \right)}^s}\lambda _D^{{\upsilon _1}}{{\left( {\Theta  - 1} \right)}^{{\upsilon _1}}}}}{{s!}}} G_{r + 2,3r + 2}^{3r,2}\left[ {\rho \left( {\Theta  - 1} \right)\left| {_{{K_2},0, - {\upsilon _1}}^{1 - {\upsilon _1},1,{K_1}}} \right.} \right]\\
& - \frac{{A\Xi }}{{\Gamma \left( {{\tau _D}} \right)}}\sum\limits_{n = 0}^{{\tau _E} - 1} {\sum\limits_{s = 0}^\infty  {\sum\limits_{t = 0}^\infty  {\frac{{{{\left( { - 1} \right)}^{s + t}}\lambda _D^{{\tau _D} + t}\lambda _E^{n + s}{{\left( {\Theta  - 1} \right)}^{{\upsilon _2}}}}}{{n!s!t!}}} } } G_{r + 2,3r + 2}^{3r,2}\left[ {\frac{{{B^r}\left( {\Theta  - 1} \right)}}{{{r^{2r}}}}\left| {_{{K_2}, - {\upsilon _2}, - n - s}^{1 - n - s,1 - {\upsilon _2},{K_1}}} \right.} \right]
\end{aligned}
\end{equation}
where ${\upsilon _1} = {\tau _D} + s$ and ${\upsilon _2} = {\tau _D} + n + s + t$.

Similarly, using of (2.24.2.1) of \cite{Prudnikov1992Vol3} leads to the following expression for $H_{22}$ as
\begin{equation}
\begin{aligned}
{H_{22}} &= \int_{\Theta  - 1}^\infty  {\int_0^{\Theta  - 1} {{F_{{\gamma _{RE}}}}\left( y \right){f_{{\gamma _{SR}}}}\left( y \right)dy} {f_{{\gamma _{RD}}}}\left( x \right)dx} \\
& = \int_{\Theta  - 1}^\infty  {{f_{{\gamma _{RD}}}}\left( x \right)dx} \int_0^{\Theta  - 1} {{F_{{\gamma _{RE}}}}\left( y \right){f_{{\gamma _{SR}}}}\left( y \right)dy} \\
& = \left( {1 - {F_{{\gamma _{RD}}}}\left( {\Theta  - 1} \right)} \right) \int_0^{\Theta  - 1} {{F_{{\gamma _{RE}}}}\left( y \right){f_{{\gamma _{SR}}}}\left( y \right)dy} \\
& = \left( {1 - {F_{{\gamma _{RD}}}}\left( {\Theta  - 1} \right)} \right) \left( {{F_{{\gamma _{SR}}}}\left( {\Theta  - 1} \right) - A\sum\limits_{n = 0}^{{\tau _E} - 1} {\frac{{\lambda _E^n{G_1}\left( {n,{\lambda _E}} \right)}}{{n!}}} } \right)
\end{aligned}
\end{equation}
where ${G_1}\left( { z_1 , z_2 } \right)$ is given by (\ref{G1fun}) in Appendix A.

After exchanging the order of the integral, we can rewrite $H_{23}$ as
\begin{equation}
\begin{aligned}
{H_{23}} &= \int_{\Theta  - 1}^\infty  {\int_{\Theta  - 1}^x {{\phi _3}\left( y \right){f_{{\gamma _{SR}}}}\left( y \right)dy} {f_{{\gamma _{RD}}}}\left( x \right)dx} \\
& = \int_{\Theta  - 1}^\infty  {{\phi _3}\left( y \right){f_{{\gamma _{SR}}}}\left( y \right)\int_y^\infty  {{f_{{\gamma _{RD}}}}\left( x \right)dx} dy} \\
& = \int_{\Theta  - 1}^\infty  {{\phi _3}\left( y \right){f_{{\gamma _{SR}}}}\left( y \right)\left( {1 - {F_{{\gamma _{RD}}}}\left( y \right)} \right)dy}.
\label{H232}
\end{aligned}
\end{equation}

By substituting (\ref{pdfsr}), (\ref{pdfrd}), and (\ref{cdfre}) into (\ref{H232}), we obtain
\begin{equation}
\begin{aligned}
{H_{23}} &= {e^{ - {\lambda _E}\left( {\frac{{1 - \Theta }}{\Theta }} \right)}}\sum\limits_{p = 0}^{{\tau _D} - 1} {\sum\limits_{n = 0}^{{\tau _E} - 1} {\sum\limits_{t = 0}^n {\frac{{{{\left( {1 - \Theta } \right)}^{n - t}}\lambda _D^p\lambda _E^n}}{{p!t!\left( {n - t} \right)!{\Theta ^n}}}} } }  {G_3}\left( {p + t - 1,{\lambda _D} + \frac{{{\lambda _E}}}{\Theta }} \right)\\
& - \sum\limits_{p = 0}^{{\tau _D} - 1} {\sum\limits_{n = 0}^{{\tau _E} - 1} {\frac{{\lambda _D^p\lambda _E^n}}{{p!n!}}{G_3}\left( {p + n - 1,{\lambda _D} + {\lambda _E}} \right)} } ,
\end{aligned}
\end{equation}
where $G_3 \left( {\alpha ,\beta } \right) = A\int_{\Theta  - 1}^\infty  {{y^\alpha }{e^{ - \beta y}}G_{1,3}^{3,0}\left[ {B{y^{\frac{1}{r}}}\left| {_{{\xi ^2},a,b}^{{\xi ^2} + 1}} \right.} \right]dy} $. With a similar method to $G_2 \left( {\alpha ,\beta } \right)$ (defined in Appendix B), we obtain the closed-form expression of $G_3 \left( {\alpha ,\beta } \right)$ as
\begin{equation}
\begin{aligned}
G_3 \left( {\alpha ,\beta } \right)  &= \frac{A\Xi }{{{\beta ^{\alpha  + 1}}}}G_{r + 1,3r}^{3r,1}\left[ {\frac{{{B^r}}}{{{r^{2r}}\beta }}\left| {_{{K_2}}^{ - \alpha ,{K_1}}} \right.} \right] \\
&- AG_1 \left( {\alpha + 1,\beta } \right).
\end{aligned}
\end{equation}

\subsection{Derivation of $\varrho$}
The PDF of ${\gamma _{eq,D}}$ can be expressed as \cite{Zedini2015IPJ}:
\begin{equation}
\begin{aligned}
{F_{\gamma _{eq,D}}}\left( \gamma  \right) &= 1 - {e^{ - {\lambda _D}\gamma }}\sum\limits_{p = 0}^{{\tau _D} - 1} {\frac{{\lambda _D^p}}{{p!}}{\gamma ^p}} + I {e^{ - {\lambda _D}\gamma }}G_{r + 1,3r + 1}^{3r,1}\left[ {\rho \gamma \left| {_{{K_2},0}^{1,{K_1}}} \right.} \right]\sum\limits_{p = 0}^{{\tau _D} - 1} {\frac{{\lambda _D^p}}{{p!}}{\gamma ^p}}.
\end{aligned}
\label{CDFmin}
\end{equation}

When considering (\ref{pdfre}) and (\ref{CDFmin}) and using (3.326.2) of \cite{Gradshteyn2007}, (11), (21) of \cite{Adamchik1990212}, we obtain
\begin{equation}
\begin{aligned}
\varrho &= \Pr \left\{ {{\gamma _{RE}} \le {\gamma _{eq,D}}} \right\}\\
& = 1 - {E_{{\gamma _{RE}}}}\left[ {\Pr \left\{ {{\gamma _{eq,D}} \le {\gamma _{RE}}\left| {{\gamma _{RE}}} \right.} \right\}} \right]\\
& =  \frac{{\lambda _E^{{\tau _E}}}}{{\Gamma \left( {{\tau _E}} \right)}}\sum\limits_{p = 0}^{{\tau _D} - 1} {\frac{{\lambda _D^p}}{{p!}}\int_0^\infty  {{x^{{\tau _E} + p - 1}}{e^{ - \left( {{\lambda _D} + {\lambda _E}} \right)x}}dx} } \\
& - \frac{{I \lambda _E^{{\tau _E}}}}{{\Gamma \left( {{\tau _E}} \right)}}\sum\limits_{p = 0}^{{\tau _D} - 1} {\frac{{\lambda _D^p}}{{p!}}\int_0^\infty  {{x^{{\tau _E} + p - 1}}{e^{ - \left( {{\lambda _D} + {\lambda _E}} \right)x}}} } G_{r + 1,3r + 1}^{3r,1}\left[ {\rho x\left| {_{{K_2},0}^{1,{K_1}}} \right.} \right]dx\\
& = \frac{{\lambda _E^{{\tau _E}}}}{{\Gamma \left( {{\tau _E}} \right)}}\sum\limits_{p = 0}^{{\tau _D} - 1} {\frac{{\lambda _D^p}}{{{{\left( {{\lambda _D} + {\lambda _E}} \right)}^{{\tau _E} + p}}p!}}} \\
& \times \left( {\Gamma \left( {{\tau _E} + p} \right) - IG_{r + 2,3r + 1}^{3r,2}\left[ {\frac{\rho }{{{\lambda _D} + {\lambda _E}}}\left| {_{{K_2},0}^{1,1 - \left( {{\tau _E} + p} \right),{K_1}}} \right.} \right]} \right).
\end{aligned}
\label{p12}
\end{equation}

\section{Asymptotic analysis of secrecy outage probability}
In this section, we analyze the asymptotic SOP in high-SNR region. We assume that ${\Omega _{SR}} = \varphi {\Omega _{RD}}$, where $\varphi $ is a constant and $\Omega _{RD}  \to \infty $.

The asymptotic CDF and PDF of ${\gamma _{RD}}$ can be given by \cite{Lei2017TVTTAS}
\begin{equation}
F_{{\gamma _{RD}}}^\infty \left( \gamma  \right) = \frac{{\phi _D^{{\tau _D}}}}{{\Omega _{RD}^{{\tau _D}}{\tau _D}!}}{\gamma ^{{\tau _D}}},
\label{cdfrdasy}
\end{equation}
\begin{equation}
f_{{\gamma _{RD}}}^\infty \left( \gamma  \right) = \frac{{\phi _D^{{\tau _D}}}}{{\Omega _{RD}^{{\tau _D}}\Gamma \left( {{\tau _D}} \right)}}{\gamma ^{{\tau _D} - 1}},
\label{pdfrdasy}
\end{equation}
respectively, where ${\phi _D} = \frac{{{m_D}d_D^\eta \left( {{\alpha _D}{N_0} + \sigma _D^2} \right)}}{{{\alpha _D}{P_t}{L_c}}}$.

The asymptotic CDF and PDF of ${\gamma _{SR}}$ can be obtained by \cite{Ansari2015VTC, Zedini2015IPJ}
\begin{equation}
F_{{\gamma _{SR}}}^\infty \left( \gamma  \right) = I \sum\limits_{k = 1}^{3r} {\frac{{{\chi _k}}}{{{K_{2,k}}\Omega _{RD}^{{K_{2,k}}}}}{\gamma ^{{K_{2,k}}}}},
\label{cdfsrasy}
\end{equation}
\begin{equation}
f_{{\gamma _{SR}}}^\infty \left( \gamma  \right) = I \sum\limits_{k = 1}^{3r} {\frac{{{\chi _k}}}{{\Omega _{RD}^{{K_{2,k}}}}}{\gamma ^{{K_{2,k}} - 1}}},
\label{pdfsrasy}
\end{equation}
respectively, where ${\chi _k} = {{{\left( {\frac{{{{\left( {hab} \right)}^2}}}{{{r^{2r}}\varphi }}} \right)}^{{K_{2,k}}}}\frac{{\prod\limits_{j = 1,j \ne k}^{3r} {\Gamma \left( {{K_{2,j}} - {K_{2,k}}} \right)} }}{{\prod\limits_{j = n + 1}^p {\Gamma \left( {{K_{1,j}} - {K_{2,k}}} \right)} }}}$.

In the following, we derive the closed-form expressions of asymptotic $H_{11}$, $H_{12}$, $H_{13}$, $H_{21}$, $H_{22}$, $H_{23}$, and $\varrho$, respectively.
\subsection{Derivation of $H_1^\infty$}
Substituting (\ref{cdfre}), (\ref{pdfrdasy}), and (\ref{pdfsrasy}) into $H_{11}$ and using (3.351.1) of \cite{Gradshteyn2007}, we have results in the following
\begin{equation}
\begin{aligned}
H_{11}^\infty & =  \int_0^{\Theta  - 1} { {\int_0^x {{F_{{\gamma _{RE}}}}\left( y \right)f_{{\gamma _{RD}}}^\infty \left( y \right)dy} } f_{{\gamma _{SR}}}^\infty \left( x \right)dx} \\
& = \frac{{I\phi _D^{{\tau _D}}}}{{\Omega _{RD}^{{\tau _D}}{\tau _D}!}}\sum\limits_{k = 1}^{3r} {\frac{{{\chi _k}}}{{\Omega _{RD}^{{K_{2,k}}}}}} \int_0^{\Theta  - 1} {{x^{{K_{2,k}} + {\tau _D} - 1}}dx}\\
& - \frac{{I\phi _D^{{\tau _D}}}}{{\Omega _{RD}^{{\tau _D}}\Gamma \left( {{\tau _D}} \right)}}\sum\limits_{k = 1}^{3r} {\frac{{{\chi _k}}}{{\Omega _{RD}^{{K_{2,k}}}}}} \sum\limits_{n = 0}^{{\tau _E} - 1} {\frac{{\lambda _E^n}}{{n!}}} \frac{{\Gamma \left( {{\tau _D} + n} \right)}}{{{\lambda _E^{{\tau _D}} + n}}}\\
& \times \int_0^{\Theta  - 1} {{x^{{K_{2,k}} - 1}}\left( {1 - {e^{ - {\lambda _E}x}}\sum\limits_{q = 0}^{{\tau _D} + n - 1} {\frac{{{\lambda _E^q}{x^q}}}{{q!}}} } \right)dx} \\
& = \frac{{I\phi _D^{{\tau _D}}}}{{\Omega _{RD}^{{\tau _D}}{\tau _D}!}}\sum\limits_{k = 1}^{3r} {\frac{{{\chi _k}{{\left( {\Theta  - 1} \right)}^{{K_{2,k}} + {\tau _D}}}}}{{\Omega _{RD}^{{K_{2,k}}}\left( {{K_{2,k}} + {\tau _D}} \right)}}} - \frac{{I\phi _D^{{\tau _D}}}}{{\Omega _{RD}^{{\tau _D}}\Gamma \left( {{\tau _D}} \right)}}\sum\limits_{k = 1}^{3r} {\frac{{{\chi _k}}}{{\Omega _{RD}^{{K_{2,k}}}}}} \sum\limits_{n = 0}^{{\tau _E} - 1} {\frac{{\Gamma \left( {{\tau _D} + n} \right)}}{{n!{\lambda _E^{{\tau _D}}}}}} \\
& \times \left( {\frac{{{{\left( {\Theta  - 1} \right)}^{{K_{2,k}}}}}}{{{K_{2,k}}}} - \sum\limits_{q = 0}^{{\tau _D} + n - 1} {\frac{{{\lambda _E^q}\Upsilon \left( {{K_{2,k}} + q,{\lambda _E}\left( {\Theta  - 1} \right)} \right)}}{{q!{{\lambda _E^{{K_{2,k}} + q}} }}}} } \right).
\label{H11asy}
\end{aligned}
\end{equation}

Similarly, we can have
\begin{equation}
\begin{aligned}
H_{12}^\infty  &= \int_{\Theta  - 1}^\infty  {\int_0^{\Theta  - 1} {{F_{{\gamma _{RE}}}}\left( y \right)f_{{\gamma _{RD}}}^\infty \left( y \right)dy} f_{{\gamma _{SR}}}^\infty \left( x \right)dx} \\
& = \int_{\Theta  - 1}^\infty  {f_{{\gamma _{SR}}}^\infty \left( x \right)dx} \left( {F_{{\gamma _{RD}}}^\infty \left( {\Theta  - 1} \right) - \frac{{\phi _D^{{\tau _D}}}}{{\Omega _D^{{\tau _D}}\Gamma \left( {{\tau _D}} \right)}}\sum\limits_{n = 0}^{{\tau _E} - 1} {\frac{{\lambda _E^n}}{{n!}}} \int_0^{\Theta  - 1} {{y^{{\tau _D} + n - 1}}{e^{ - {\lambda _E}y}}dy} } \right)\\
& = \left( {1 - F_{{\gamma _{SR}}}^\infty \left( {\Theta  - 1} \right)} \right)\left( {F_{{\gamma _{RD}}}^\infty \left( {\Theta  - 1} \right) - \frac{{\phi _D^{{\tau _D}}}}{{\Omega _D^{{\tau _D}}\Gamma \left( {{\tau _D}} \right)}}\sum\limits_{n = 0}^{{\tau _E} - 1} {\frac{{\Upsilon \left( {{\tau _D} + n,{\lambda _E}\left( {\Theta  - 1} \right)} \right)}}{{n!\lambda _E^{{\tau _D}}}}} } \right).
\label{H12asy}
\end{aligned}
\end{equation}
\begin{equation}
\begin{aligned}
H_{13}^\infty &= \int_{\Theta  - 1}^\infty  {\int_{\Theta  - 1}^x {{\phi _3}\left( y \right)f_{{\gamma _{RD}}}^\infty \left( y \right)dy} f_{{\gamma _{SR}}}^\infty \left( x \right)dx} \\
& = {e^{ - {\lambda _E}\left( {\frac{{1 - \Theta }}{\Theta }} \right)}}\sum\limits_{n = 0}^{{\tau _E} - 1} {\sum\limits_{q = 0}^n {\frac{{\lambda _E^n{{\left( {1 - \Theta } \right)}^{n - q}}}}{{{\Theta ^n}q!\left( {n - q} \right)!}}{\psi _1}\left( {q,\frac{{{\lambda _E}}}{\Theta }} \right)} } - \sum\limits_{n = 0}^{{\tau _E} - 1} {\frac{{\lambda _E^n}}{{n!}}{\psi _1}\left( {n,{\lambda _E}} \right)},
\label{H13asy}
\end{aligned}
\end{equation}
where ${\psi _1}\left( {{c_1},{c_2}} \right) = \int_{\Theta  - 1}^\infty  {\int_{\Theta  - 1}^x {{y^{{c_1}}}{e^{ - {c_2}y}}f_{{\gamma _{RD}}}^\infty \left( y \right)dy} f_{{\gamma _{SR}}}^\infty \left( x \right)dx} $.

By using (3.326.2), (3.351.2), and (8.352.6) of \cite{Gradshteyn2007}, we obtain
\begin{equation}
\begin{aligned}
{\psi _1}\left( {{c_1},{c_2}} \right) &= \int_{\Theta  - 1}^\infty  {\int_{\Theta  - 1}^x {{y^{{c_1}}}{e^{ - {c_2}y}}f_{{\gamma _{RD}}}^\infty \left( y \right)dy} f_{{\gamma _{SR}}}^\infty \left( x \right)dx}  \\
& = \frac{{\phi _D^{{\tau _D}}\Gamma \left( {{\tau _D} + {c_1}} \right)}}{{c_2^{{\tau _D} + {c_1}}\Omega _{RD}^{{\tau _D}}\Gamma \left( {{\tau _D}} \right)}}\left( {1 - F_{{\gamma _{SR}}}^\infty \left( {\Theta  - 1} \right) - I\sum\limits_{n = 0}^{{\tau _D} + {c_1} - 1} {\sum\limits_{k = 1}^{3r} {\frac{{{\chi _k}\Gamma \left( {{K_{2,k}} + n,{c_2}\left( {\Theta  - 1} \right)} \right)}}{{n!\Omega _{RD}^{{K_{2,k}}}c_2^{{K_{2,k}}}}}} } } \right)\\
&  - \frac{{\phi _D^{{\tau _D}}\Upsilon \left( {{\tau _D} + {c_1},{c_2}\left( {\Theta  - 1} \right)} \right)}}{{c_2^{{\tau _D} + {c_1}}\Omega _{RD}^{{\tau _D}}\Gamma \left( {{\tau _D}} \right)}} \left( {1 - F_{{\gamma _{SR}}}^\infty \left( {\Theta  - 1} \right)} \right).
\end{aligned}
\end{equation}

\subsection{Derivation of $H_2^\infty$}
By substituting (\ref{cdfre}), (\ref{pdfrdasy}), and (\ref{pdfsrasy}) into $H_{21}$ and using (\ref{lowerimcomplete}) of Appendix A, and (07.34.21.0084.01) of \cite{Wolfram2001}, we obtain
\begin{equation}
\begin{aligned}
H_{21}^\infty  &= \int_0^{\Theta  - 1} { {\int_0^x {{F_{{\gamma _{RE}}}}\left( y \right)f_{{\gamma _{SR}}}^\infty \left( y \right)dy} }f_{{\gamma _{RD}}}^\infty \left( x \right)dx} \\
& = \int_0^{\Theta  - 1} {F_{{\gamma _{SR}}}^\infty \left( x \right)f_{{\gamma _{RD}}}^\infty \left( x \right)dx} \\
& - I \sum\limits_{k = 1}^{3r} {\frac{{{\chi _k}}}{{\Omega _{RD}^{{K_{2,k}}}}}} \sum\limits_{n = 0}^{{\tau _E} - 1} {\frac{1}{{n!{{\lambda _E^{{K_{2,k}}}}}}}} \int_0^{\Theta  - 1} {\Upsilon \left( {{K_{2,k}} + n,{\lambda _E}x} \right)f_{{\gamma _{RD}}}^\infty \left( x \right)dx} \\
& = \frac{{I\phi _D^{{\tau _D}}}}{{\Omega _{RD}^{{\tau _D}}\Gamma \left( {{\tau _D}} \right)}}\sum\limits_{k = 1}^{3r} {\frac{{{\chi _k}}}{{{K_{2,k}}\Omega _{RD}^{{K_{2,k}}}}}\int_0^{\Theta  - 1} {{x^{{K_{2,k}} + {\tau _D} - 1}}dx} } \\
& - \frac{{I\phi _D^{{\tau _D}}}}{{\Omega _{RD}^{{\tau _D}}\Gamma \left( {{\tau _D}} \right)}}\sum\limits_{k = 1}^{3r} {\sum\limits_{n = 0}^{{\tau _E} - 1} {\frac{{{\chi _k}}}{{n!\Omega _{RD}^{{K_{2,k}}}{{\lambda _E^{{K_{2,k}}}}}}}}} \int_0^{\Theta  - 1} {\Upsilon \left( {{K_{2,k}} + n,{\lambda _E}x} \right){x^{{\tau _D} - 1}}dx} \\
& = \frac{{I\phi _D^{{\tau _D}}}}{{\Omega _{RD}^{{\tau _D}}\Gamma \left( {{\tau _D}} \right)}}\sum\limits_{k = 1}^{3r} {\frac{{{\chi _k}{{\left( {\Theta  - 1} \right)}^{{K_{2,k}} + {\tau _D}}}}}{{{K_{2,k}}\Omega _{RD}^{{K_{2,k}}}\left( {{K_{2,k}} + {\tau _D}} \right)}}} \\
& - \frac{{{{\left( {\Theta  - 1} \right)}^{{\tau _D}}}I {\phi _D^{{\tau _D}}}}}{{\Omega _{RD}^{{\tau _D}}\Gamma \left( {{\tau _D}} \right)}}\sum\limits_{k = 1}^{3r} {\sum\limits_{n = 0}^{{\tau _E} - 1} {\frac{{{\chi _k}}}{{\Omega _{RD}^{{K_{2,k}}}n!{{\lambda _E^{{K_{2,k}}}}}}}}} G_{2,3}^{1,2}\left[ {{\lambda _E}\left( {\Theta  - 1} \right)\left| {_{{K_{2,k}} + n,0, - {\tau _D}}^{1 - {\tau _D},1}} \right.} \right].
\label{H21asy}
\end{aligned}
\end{equation}

Similarly, we have
\begin{equation}
\begin{aligned}
H_{22}^\infty  &= \int_{\Theta  - 1}^\infty  {\int_0^{\Theta  - 1} {{F_{{\gamma _{RE}}}}\left( y \right)f_{{\gamma _{SR}}}^\infty \left( y \right)dy} f_{{\gamma _{RD}}}^\infty \left( x \right)dx} \\
& = \int_{\Theta  - 1}^\infty  {f_{{\gamma _{RD}}}^\infty \left( x \right)dx} \int_0^{\Theta  - 1} {{F_{{\gamma _{RE}}}}\left( y \right)f_{{\gamma _{SR}}}^\infty \left( y \right)dy} \\
& = \left( {1 - F_{{\gamma _{RD}}}^\infty \left( {\Theta  - 1} \right)} \right)\left( {F_{{\gamma _{SR}}}^\infty \left( {\Theta  - 1} \right) - I\sum\limits_{n = 0}^{{\tau _E} - 1} {\sum\limits_{k = 1}^{3r} {\frac{{{\chi _k}\Upsilon \left( {{K_{2,k}} + n,{\lambda _E}\left( {\Theta  - 1} \right)} \right)}}{{n!\Omega _{RD}^{{K_{2,k}}}{\lambda _E}^{{K_{2,k}}}}}} } } \right).
\label{H22asy}
\end{aligned}
\end{equation}

By adopting (\ref{cdfre}), (\ref{H232}), (\ref{cdfrdasy}), and (\ref{pdfsrasy}), and after some algebraic manipulations, we obtain
\begin{equation}
\begin{aligned}
H_{23}^\infty  &= \int_{\Theta  - 1}^\infty  {\left( {1 - F_{{\gamma _{RD}}}^\infty \left( y \right)} \right){\phi _3}\left( y \right)f_{{\gamma _{SR}}}^\infty \left( y \right)dy} \\
& = {e^{ - {\lambda _E}\left( {\frac{{1 - \Theta }}{\Theta }} \right)}}\sum\limits_{n = 0}^{{\tau _E} - 1} {\sum\limits_{q = 0}^n {\frac{{\lambda _E^n{{\left( {1 - \Theta } \right)}^{n - q}}}}{{{\Theta ^n}q!\left( {n - q} \right)!}}} } \\
& \times \left( {{\psi _2}\left( {q,\frac{{{\lambda _E}}}{\Theta }} \right) - \frac{{\phi _D^{{\tau _D}}}}{{\Omega _{RD}^{{\tau _D}}{\tau _D}!}}{\psi _2}\left( {{\tau _D} + q,\frac{{{\lambda _E}}}{\Theta }} \right)} \right)\\
& + \sum\limits_{n = 0}^{{\tau _E} - 1} {\frac{{\lambda _E^n}}{{n!}}\left( {\frac{{\phi _D^{{\tau _D}}}}{{\Omega _{RD}^{{\tau _D}}{\tau _D}!}}{\psi _2}\left( {{\tau _D} + n,{\lambda _E}} \right) - {\psi _2}\left( {n,{\lambda _E}} \right)} \right)},
\label{H23asy}
\end{aligned}
\end{equation}
where ${\psi _2}\left( {{c_1},{c_2}} \right) = \int_{\Theta  - 1}^\infty  {{\gamma ^{{c_1}}}{e^{ - {c_2}y}}f_{{\gamma _{SR}}}^\infty \left( y \right)dy}$.

By placing ${f_{{\gamma _{SR}}}^\infty \left( y \right)}$ into ${\psi _2}\left( {{c_1},{c_2}} \right)$ and using (3.351.2) of \cite{Gradshteyn2007}, one can have
\begin{equation}
\begin{aligned}
{\psi _2}\left( {{c_1},{c_2}} \right) &= \int_{\Theta  - 1}^\infty  {{\gamma ^{{c_1}}}{e^{ - {c_2}y}}f_{{\gamma _{SR}}}^\infty \left( y \right)dy} \\
& = I\sum\limits_{k = 1}^{3r} {\frac{{{\chi _k}}}{{\Omega _{RD}^{{K_{2,k}}}}}} \int_{\Theta  - 1}^\infty  {{y^{{K_{2,k}} + {c_1} - 1}}{e^{ - {c_2}y}}dy} \\
& = I\sum\limits_{k = 1}^{3r} {\frac{{{\chi _k}\Gamma \left( {{K_{2,k}} + {c_1},{c_2}\left( {\Theta  - 1} \right)} \right)}}{{\Omega _{RD}^{{K_{2,k}}}c_2^{{K_{2,k}} + {c_1}}}}},
\end{aligned}
\end{equation}
where $\Gamma \left( { \cdot , \cdot } \right)$ is the upper incomplete Gamma function, defined by (8.350.2) of \cite{Gradshteyn2007}.

\subsection{Derivation of $\varrho^\infty$}
The asymptotic CDF of ${\gamma _{eq,D}}$ can be expressed as \cite{Leihj2017PJ}
\begin{equation}
F_{{\gamma _{eq,D}}}^\infty \left( x \right) = \frac{{\phi _D^{{\tau _D}}{x^{{\tau _D}}}}}{{\Omega _{RD}^{{\tau _D}}{\tau _D}!}} + I \sum\limits_{k = 1}^{3r} {\frac{{{\chi _k}{x^{{K_{2,k}}}}}}{{{K_{2,k}}\Omega _{RD}^{{K_{2,k}}}}}} - \frac{{I \phi _D^{{\tau _D}}}}{{\Omega _{RD}^{{\tau _D}}{\tau _D}!}}\sum\limits_{k = 1}^{3r} {\frac{{{\chi _k}{x^{{K_{2,k}} + {\tau _D}}}}}{{{K_{2,k}}\Omega _{RD}^{{K_{2,k}}}}}}.
\label{CDFminasy}
\end{equation}

By substituting (\ref{pdfre}) and (\ref{CDFminasy}) into (\ref{p12}) and using (3.326.2) of \cite{Gradshteyn2007}, it deduces
\begin{equation}
\begin{aligned}
\varrho^\infty &= 1 -  \frac{{\phi _D^{{\tau _D}}\Gamma \left( {{\tau _E} + {\tau _D}} \right)}}{{\Omega _{RD}^{{\tau _D}}{\tau _D}!\Gamma \left( {{\tau _E}} \right){\lambda _E}^{{\tau _D}}}} - \frac{{I \lambda _E^{{\tau _E}}}}{{\Gamma \left( {{\tau _E}} \right)}}\sum\limits_{k = 1}^{3r} {\frac{{{\chi _k}\Gamma \left( {{K_{2,k}} + {\tau _E}} \right)}}{{{K_{2,k}}\Omega _{RD}^{{K_{2,k}}}{\lambda _E}^{{K_{2,k}} + {\tau _E}}}}} \\
& + \frac{{I \phi _D^{{\tau _D}}}}{{\Omega _{RD}^{{\tau _D}}{\tau _D}!\Gamma \left( {{\tau _E}} \right)}}\sum\limits_{k = 1}^{3r} {\frac{{{\chi _k}\Gamma \left( {{K_{2,k}} + {\tau _D} + {\tau _E}} \right)}}{{{K_{2,k}}\Omega _{RD}^{{K_{2,k}}}{\lambda _E}^{{K_{2,k}} + {\tau _D}}}}}.
\label{P12asy}
\end{aligned}
\end{equation}

The SDO is expressed as \cite{Lei2017TVTTAS}
\begin{equation}
{G_d} =  - \mathop {\lim }\limits_{{\Omega _{RD}} \to \infty } \frac{{\ln P_{out}^\infty }}{{\ln {\Omega _{RD}}}},
\label{Gd}
\end{equation}
where ${P_{out}^\infty }$ denotes the asymptotic SOP.

Observing the aforementioned results, we find that $H_{12}^\infty$, $H_{13}^\infty$, $H_{22}^\infty$, $H_{23}^\infty$, and $\varrho^\infty$ are the dominant terms for SDO in the higher SNR region. We can then easily achieve secrecy diversity order as
\begin{equation}
{G_d} = \min \left\{ {{\tau _D},\frac{{{\xi ^2}}}{r},\frac{a}{r},\frac{b}{r}} \right\}.
\label{Gd2}
\end{equation}

\textcolor[rgb]{0.00,0.07,1.00}{\emph{\textbf{Remark 5:}} From (\ref{Gd2}), one can observe that the SDO depends on the fading parameter of the $R$-$D$ link ($m_D$), the number of the antennas at $D$ ($N_D$), the fading parameters $\left( {a,b} \right)$, the detection type ($r$), and the pointing error parameter ($\xi$) of the FSO link.}

\textcolor[rgb]{0.00,0.07,1.00}{\emph{\textbf{Remark 6:}} When $m_D$ or/and $N_D$ increase, which means the SNR at $D$ increases, the SDO will become better, which can be easily understandable.}

\textcolor[rgb]{0.00,0.07,1.00}{\emph{\textbf{Remark 7:}} For large $\xi$ or/and $\left( {a,b} \right)$, the SDO will get better since the SNR at $R$ get better. This can be explained by the results in \cite{lei2017CL}. With the same reason, the SDO with HD ($r = 1$) is outperforms that of IM/DD ($r = 2$).}

\textcolor[rgb]{0.00,0.07,1.00}{\emph{\textbf{Remark 8:}} An interesting result can be reached from (\ref{Gd2}) that the SDO of mixed RF-FSO systems is independent of the parameter of EH ($\alpha$), the path-loss exponent ($\eta$), the fading parameter of the eavesdropping channels ($m_E$), and the number of antennas at $E$ ($N_E$).}

\section{Numerical results}
In this section, we represent our results with figures to better clarify the SOP of the DL mixed RF-FSO SWIPT systems.
Additionally, we analyze the impacts of the following parameters on the security of DL mixed RF-FSO systems: the number of the antennas at the legitimate destination, fading parameter of RF and FSO links, pointing error, and type of detection.
In these figures, we set ${N_0} = \sigma _D^2 = \sigma _E^2 = 1$, ${P_t} = 30\,\,{\rm{dBm}}$, ${d_D} = {d_E} = 10 $ m, ${L_c} = 3.597 \times {10^{ - 2}}$, $N_E = 2$, ${\alpha _D} = {\alpha _E} = \alpha $, ${m _D} = {m _E} = m$, and $R_ s = 0.01$ nat/s/Hz.
\begin{figure}[!t]
\centering
\includegraphics[width = 4 in]{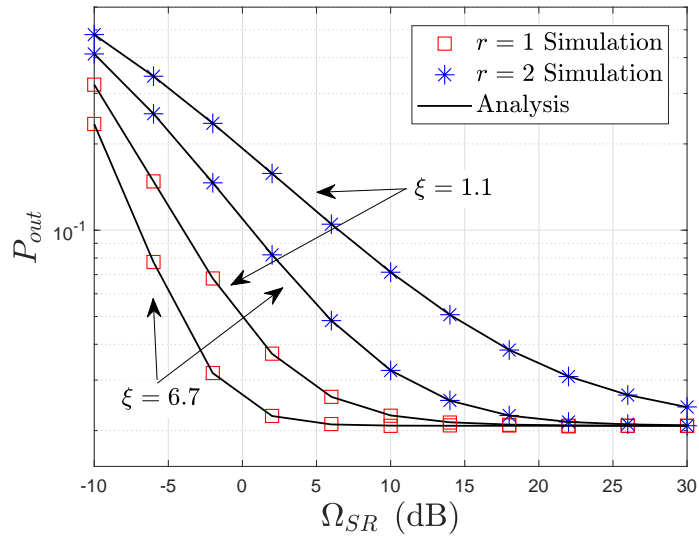}
\caption{SOP versus ${\Omega _{SR}}$  with $N_D = 3$, $\eta = 3.0$, $m = 2$, $a = 2.902$, $b = 2.51$, $\alpha = 0.5$, ${\Omega _{RD}} = 5\,{\rm{dB}}$, and ${\Omega _{RE}} = 0\,{\rm{dB}}$.}
\label{fig2}
\end{figure}
\begin{figure}[!t]
\centering
\includegraphics[width = 4 in]{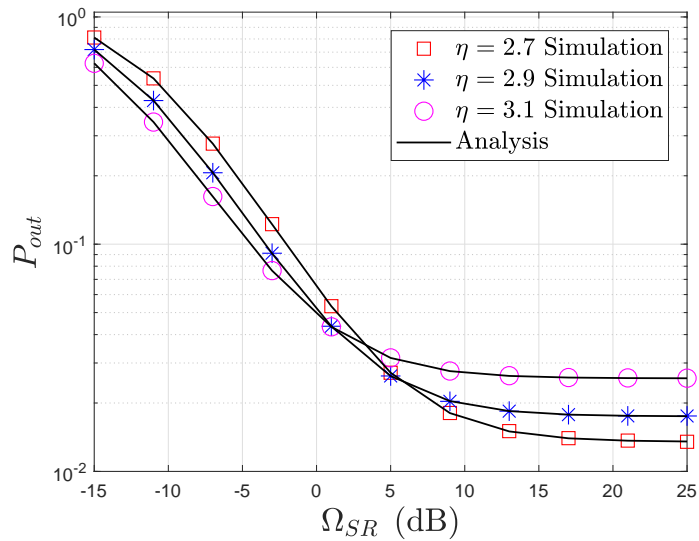}
\caption{SOP versus ${\Omega _{SR}}$ with $N_D = 3$, $m = 2$, $r = 1$, $\xi  = 1.1$, $\alpha = 0.5$, ${\Omega _{RD}} = 5\,{\rm{dB}}$, and ${\Omega _{RE}} = 0\,{\rm{dB}}$.}
\label{fig3}
\end{figure}
\begin{figure}[!t]
\centering
\includegraphics[width = 4 in]{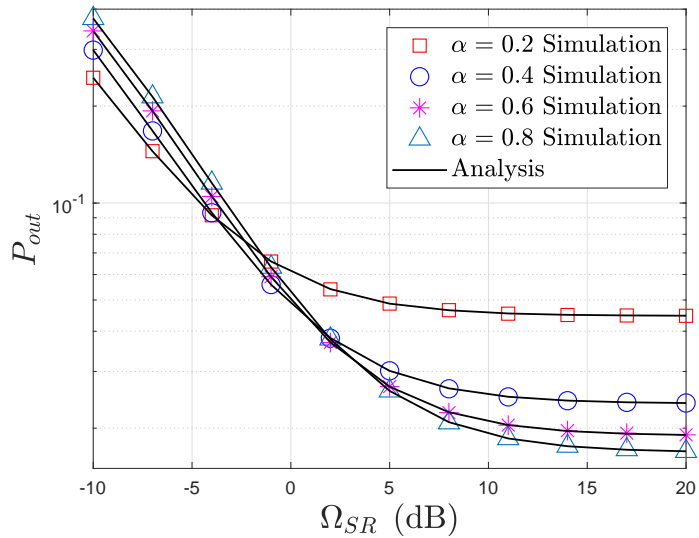}
\caption{SOP versus ${\Omega _{SR}}$ with $N_D = 3$, $m = 2$, $\eta = 3.0$, $r = 1$, $\xi  = 1.1$, ${\Omega _{RD}} = 5\,{\rm{dB}}$, and ${\Omega _{RE}} = 0\,{\rm{dB}}$.}
\label{fig4}
\end{figure}
\begin{figure}[!t]
\centering
\includegraphics[width = 4 in]{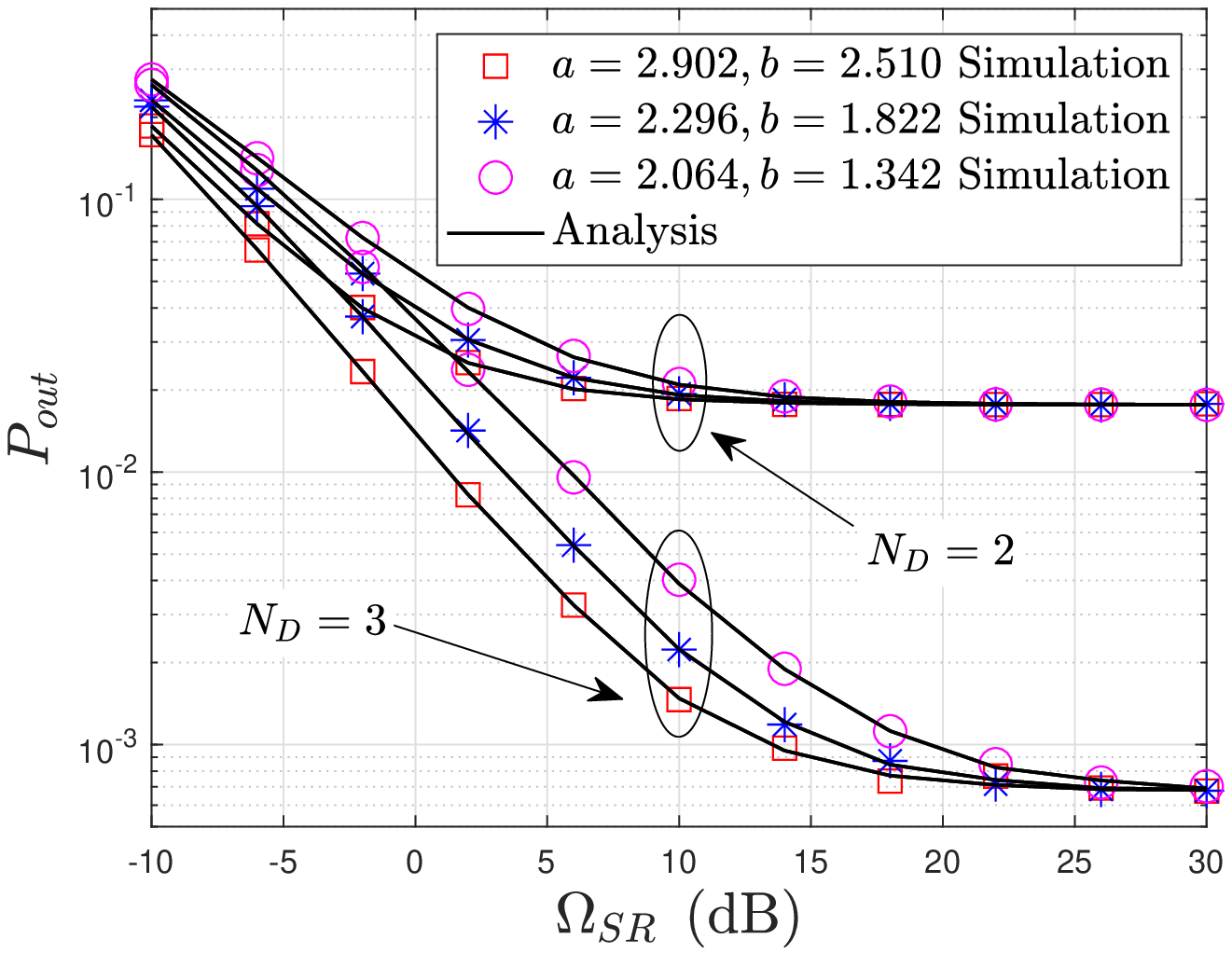}
\caption{SOP versus ${\Omega _{SR}}$ with $m = 2$, $\eta = 3.0$, $r = 1$, $\xi  = 1.1$, $\alpha = 0.5$, ${\Omega _{RD}} = 5\,{\rm{dB}}$, and ${\Omega _{RE}} = 0\,{\rm{dB}}$.}
\label{fig5}
\end{figure}

Figs. \ref{fig2} - \ref{fig5} show the SOP versus ${\Omega _{SR}}$ for different values of  $r$, $\xi$, $N_D$, $\eta$, and $\alpha$. These figures clearly show that the secrecy performance is enhanced with increasing ${\Omega _{SR}}$, as the SNR at the relay node is improved. By varying $r$ and keeping $\xi$ fixed in Fig. \ref{fig2}, the HD detection method can lead to better secrecy performance than IM/DD method. Moreover, the SOP with lower $\xi$ is higher than that with larger $\xi$. The reason for this is that the SNR obtained with the HD method is higher than that of IM/DD, and a larger $\xi$ means higher pointing accuracy. Finally, SOP exhibits a floor because the secrecy capacity will become a constant, as testified in \cite{lei2017CL}.

\textcolor[rgb]{0.00,0.07,1.00}{As shown in Figs. \ref{fig3} and \ref{fig4}, the SOPs with lower $\eta$ or higher $\alpha$ values outperform those with higher $\eta$ or lower $\alpha$ value in higher ${\Omega _{SR}}$ regions. This is because a lower $\eta$ represents a weaker path-loss for the RF signals, and a higher $\alpha$ means more power is allocated to decoding the information, resulting in a higher power at target destinations.}

Fig. \ref{fig5} represents the SOP for different values $\left( {a,b} \right)$ and $N_D$. We can see that a larger $N_D$ will result in a smaller SOP, meaning better secrecy performance, since a larger $N_D$ leads to more diversity gains. One can also see that the SOP with the weak turbulence ($a = 2.902$, $b = 2.510$) is lower than that with strong turbulence ($a = 2.064$, $b = 1.342$). The reason is the same as that for the previous findings in Fig. \ref{fig2}.

\begin{figure}[!t]
\centering
\includegraphics[width = 4 in]{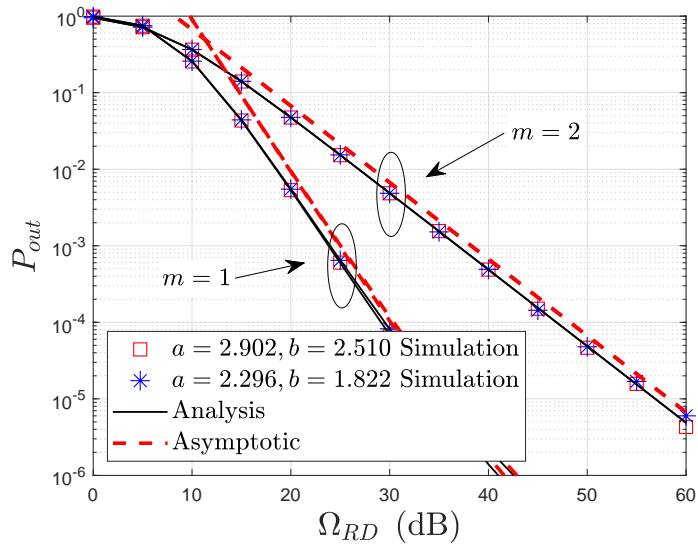}
\caption{SOP versus ${\Omega _{RD}}$ with $N_D = 1$, $\eta = 3.0$, $r = 1$, $\xi  = 1.1$, $\alpha = 0.5$, and ${\Omega _{RE}} = 3\,{\rm{dB}}$.}
\label{fig6}
\end{figure}

\begin{figure}[!t]
\centering
\includegraphics[width = 4 in]{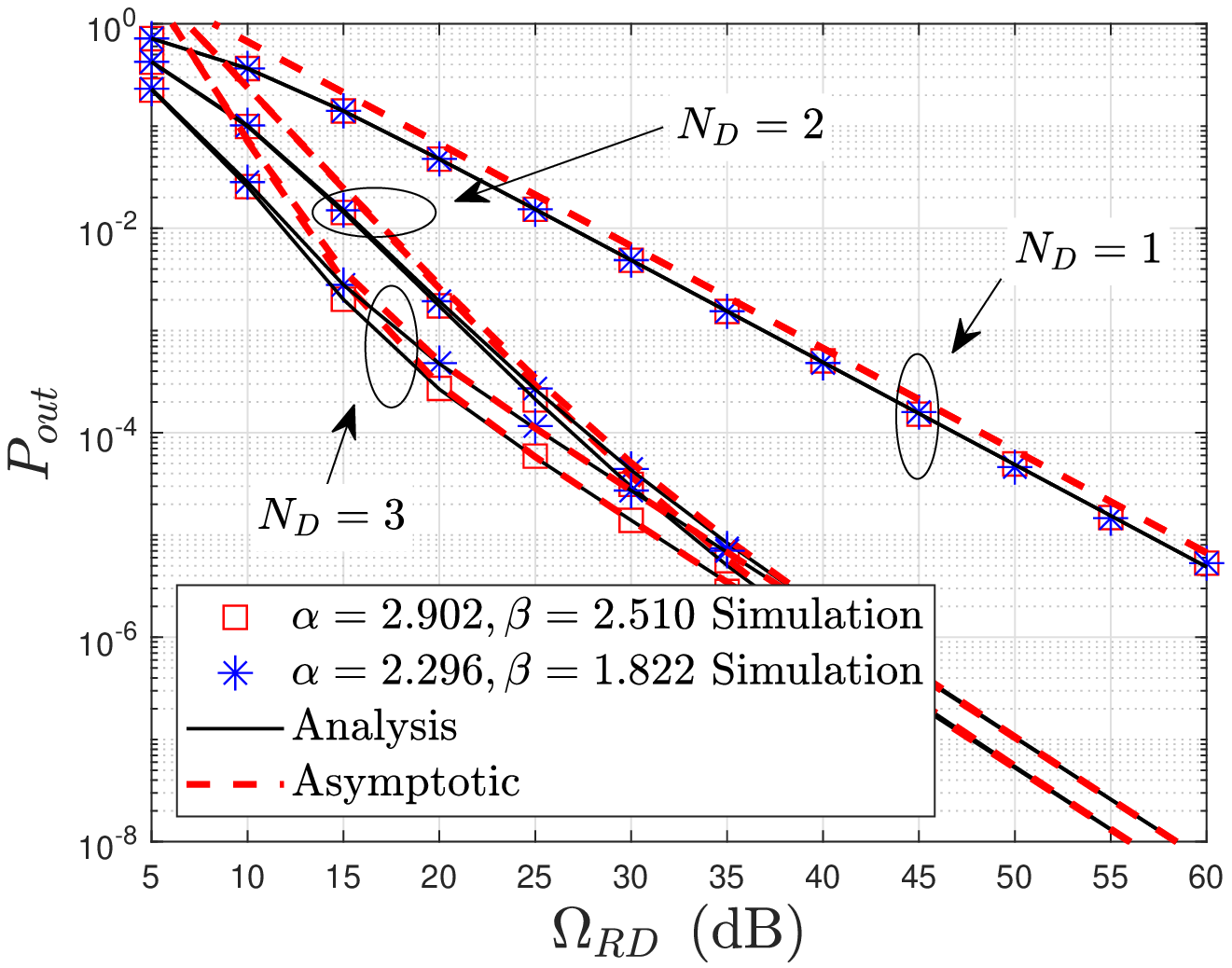}
\caption{SOP versus ${\Omega _{RD}}$ with $m = 1$, $\eta = 3.0$, $r = 1$, $\xi  = 1.1$, $\alpha = 0.5$, and ${\Omega _{RE}} = 3\,{\rm{dB}}$.}
\label{fig7}
\end{figure}
\begin{figure}[!t]
\centering
\includegraphics[width = 4 in]{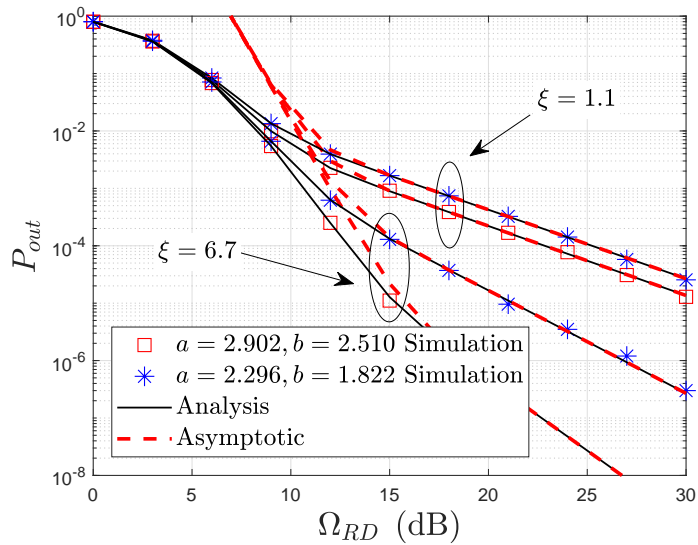}
\caption{SOP versus ${\Omega _{RD}}$ with $N_D = 3$, $m = 2$, $\eta = 3.0$, $N_E = 2$, $r = 1$, $\alpha = 0.5$, and ${\Omega _{RE}} = 3\,{\rm{dB}}$.}
\label{fig8}
\end{figure}
\begin{figure}[!t]
\centering
\includegraphics[width = 4 in]{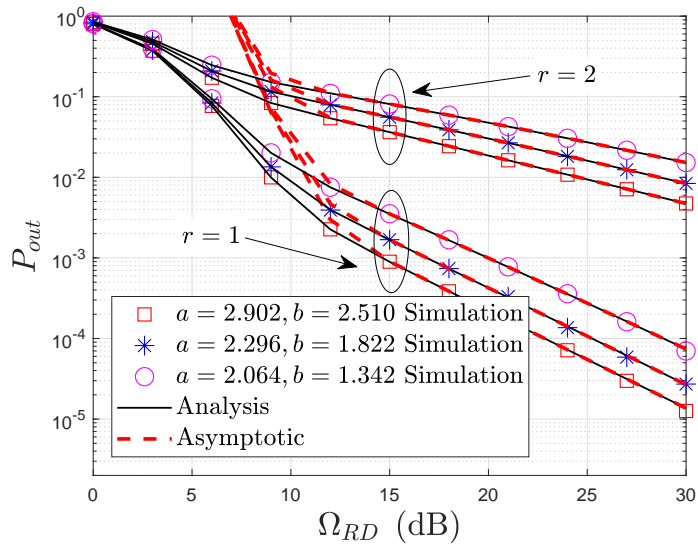}
\caption{SOP versus ${\Omega _{RD}}$ with $N_D = 3$, $m = 2$,  $\eta = 3.0$, $\xi  = 1.1$, $\alpha = 0.5$, and ${\Omega _{RE}} = 3\,{\rm{dB}}$.}
\label{fig9}
\end{figure}

For the results presented in Figs. \ref{fig6}-\ref{fig9}, we set $\varphi = 1$. In this figures, we see that the asymptotic SOP tightly approximates the exact results at high-${\Omega _{RD}}$ regime. Additionally, the SDO is subject to the fading parameters of the RF link (Fig. \ref{fig6}), the number of the antennas at $D$ (Fig. \ref{fig7}), the fading parameters of FSO link (Figs. \ref{fig6}-\ref{fig9}), the pointing errors (Fig. \ref{fig8}), and the detection technology (Fig. \ref{fig8}). These results are in agreement with those obtained using (44).

\section{Conclusion}
In this work, we analyzed the secrecy performance of DL mixed RF-FSO SWIPT systems and derived the closed-form expressions for the exact and asymptotic SOPs. Simulation and numerical results show that  the HD detection method can lead to better secrecy performance compared to IM/DD method.
The fading parameter of relay-destination link, the number of the destination's antennas, the fading parameters, the detection type, and the pointing error parameter of the FSO link will influence the SDO of DL mixed RF-FSO SWIPT systems.
Our results can be used in practical mixed RF-FSO systems design, in which security issue is considered.
\textcolor[rgb]{0.00,0.07,1.00}{The interesting topic for future work is investigating the secrecy performance of the mixed RF-FSO systems with multiple relays (such as relay selection) and multiple users (such as user scheduling) and design the cooperative jamming scheme with artificial noise to improve the secrecy performance of mixed RF- FSO systems.}

\begin{appendices}
\section{}
Using (8.352.1) of \cite{Gradshteyn2007}, we rewrite $G_0 \left( {\alpha ,\beta } \right)$ as
\begin{equation}
\begin{aligned}
G_0 \left( {\alpha ,\beta } \right) &= \int_0^{\Theta  - 1} {\Upsilon \left( {\alpha ,\beta x} \right){f_{{\gamma _{SR}}}}\left( x \right)dx} \\
& = A \Gamma \left( \alpha  \right)\int_0^{\Theta  - 1} {{x^{ - 1}}\left( {1 - e^ { - \beta x} \sum\limits_{t = 0}^{\alpha  - 1} {\frac{{{\beta ^t}{x^t}}}{{t!}}} } \right)} G_{1,3}^{3,0}\left[ {B{x^{\frac{1}{r}}}\left| {_{{\xi ^2},a,b}^{{\xi ^2} + 1}} \right.} \right]dx \\
& = A\Gamma \left( \alpha  \right)\int_0^{\Theta  - 1} {{x^{ - 1}}G_{1,3}^{3,0}\left[ {B{x^{\frac{1}{r}}}\left| {_{{\xi ^2},a,b}^{{\xi ^2} + 1}} \right.} \right]dx} \\
& - A\Gamma \left( \alpha  \right)\sum\limits_{t = 0}^{\alpha  - 1} {\frac{{{\beta ^t}}}{{t!}}}  \int_0^{\Theta  - 1} {{x^{t - 1}}\exp \left( { - \beta x} \right) G_{1,3}^{3,0}\left[ {B{x^{\frac{1}{r}}}\left| {_{{\xi ^2},a,b}^{{\xi ^2} + 1}} \right.} \right]dx} \\
& = A \Gamma \left( \alpha  \right)\left({G_1}\left( {0,0} \right) - \sum\limits_{t = 0}^{\alpha  - 1} {\frac{{{\beta ^t}{G_1}\left( {t,\beta } \right)}}{{t!}}}\right),
\label{faifun}
\end{aligned}
\end{equation}
where ${G_1}\left( {{z_1},{z_2}} \right) = \int_0^{\Theta  - 1} {{x^{{z_1} - 1}}{e^{ - {z_2}x}}G_{1,3}^{3,0}\left[ {B{x^{\frac{1}{r}}}\left| {_{{\xi ^2},a,b}^{{\xi ^2} + 1}} \right.} \right]dx} $.

Then utilizing ${e^{ - x}} = \sum\limits_{s = 0}^\infty  {\frac{{{{\left( { - x} \right)}^s}}}{{s!}}} $ and (07.34.21.0084.01) of \cite{Wolfram2001}, we obtain \footnote{Note that the infinity summation is using here, but it converges quickly, which is verified by simulation results in Section V and many literatures, such as \cite{ZhangC2017Access, FengJ2017OC}.}
\begin{equation}
\begin{aligned}
{G_1}\left( {{z_1},{z_2}} \right) &= \int_0^{\Theta  - 1} {{x^{{z_1} - 1}}{e^{ - {z_2}x}}G_{1,3}^{3,0}\left[ {B{x^{\frac{1}{r}}}\left| {_{{\xi ^2},a,b}^{{\xi ^2} + 1}} \right.} \right]dx} \\
& = \sum\limits_{s = 0}^\infty  {\frac{{{{\left( { - {z_2}} \right)}^s}}}{{s!}}} \int_0^{\Theta  - 1} {{x^{{z_1} + s - 1}}G_{1,3}^{3,0}\left[ {B{x^{\frac{1}{r}}}\left| {_{{\xi ^2},a,b}^{{\xi ^2} + 1}} \right.} \right]dx} \\
& = \Xi \sum\limits_{s = 0}^\infty  {\frac{{{{\left( { - {z_2}} \right)}^s}{{\left( {\Theta  - 1} \right)}^{{z_1} + s}}}}{{s!}}} G_{r + 1,3r + 1}^{3r,1}\left[ {\frac{{{B^r}\left( {\Theta  - 1} \right)}}{{{r^{2r}}}}\left| {_{{K_2}, - {z_1} - s}^{1 - {z_1} - s,{K_1}}} \right.} \right],
\label{G1fun}
\end{aligned}
\end{equation}
where $\Xi  = \frac{{{r^{a + b - 1}}}}{{{{\left( {2\pi } \right)}^{r - 1}}}}$.
\section{}
By using (9) of \cite{lei2015CL}, (8.311.1), and (9.31.5) of \cite{Gradshteyn2007}, we have
\begin{equation}
\Upsilon \left( {a,z} \right) = G_{1,2}^{1,1}\left[ {z\left| {_{a,0}^1} \right.} \right].
\label{lowerimcomplete}
\end{equation}
And by using (21) in \cite{Adamchik1990212}, we finally obtain
\begin{equation}
\begin{aligned}
G_2 \left( {\alpha ,\beta } \right) &= A\int_{\Theta  - 1}^\infty  {{x^{ - 1}}\Upsilon \left( {\alpha ,\beta x} \right)G_{1,3}^{3,0}\left[ {B{x^{\frac{1}{r}}}\left| {_{{\xi ^2},a,b}^{{\xi ^2} + 1}} \right.} \right]dx} \\
&  = A\int_0^\infty  {{x^{ - 1}}G_{1,2}^{1,1}\left[ {\beta x\left| {_{\alpha ,0}^1} \right.} \right]G_{1,3}^{3,0}\left[ {B{x^{\frac{1}{r}}}\left| {_{{\xi ^2},a,b}^{{\xi ^2} + 1}} \right.} \right]dx} \\
& - A\int_0^{\Theta - 1} {{x^{ - 1}}\Upsilon \left( {\alpha ,\beta x} \right)G_{1,3}^{3,0}\left[ {B{x^{\frac{1}{r}}}\left| {_{{\xi ^2},a,b}^{{\xi ^2} + 1}} \right.} \right]dx}  \\
& = A\Xi G_{r + 2,3r + 1}^{3r + 1,1}\left[ {\frac{{{B^r}}}{{{r^{2r}}\beta }}\left| {_{{K_2},0}^{1 - \alpha ,1,{K_1}}} \right.} \right] - AG_0 \left( {\alpha ,\beta } \right).
\end{aligned}
\end{equation}

\end{appendices}

\end{document}